\documentclass[
 reprint,
superscriptaddress,
 amsmath,amssymb,
 aps,
pra,
aip,
rsi,]{revtex4-1}
\usepackage{color}
\usepackage{xcolor}
\usepackage{graphicx}% Include figure files
\usepackage{caption}
\usepackage{subcaption}
\usepackage{dcolumn}% Align table columns on decimal point
\usepackage{bm}% bold math
\usepackage{braket}% braket quantum symbols
\usepackage{float}% fix figure position
\usepackage{multirow}
\usepackage{hhline}
\usepackage{mathtools}
\usepackage{qcircuit}
\usepackage{lipsum}

\definecolor{xblue}{HTML}{3378cc}
\definecolor{zorange}{HTML}{FF7F00}

\begin{document}

%\preprint{APS/123-QED}

\title{Logical Performance of 9 Qubit Compass Codes in Ion Traps with Crosstalk Errors}
\author{Dripto M. Debroy}
% \thanks{These two authors contributed equally}
\email{dripto@phy.duke.edu; contributed equally}

\affiliation{Department of Physics, Duke University, Durham, NC 27708, USA}
\author{Muyuan Li}
% \thanks{These two authors contributed equally}
\email{mli97@gatech.edu; contributed equally}
\affiliation{School of Computational Science and Engineering,
 Georgia Institute of Technology, Atlanta, Georgia 30332, USA}
 \author{Shilin Huang}
\affiliation{Department of Electrical and Computer Engineering, Duke University, Durham, NC 27708, USA}
\author{Kenneth R. Brown}
\email{ken.brown@duke.edu}
 \affiliation{Department of Physics, Duke University, Durham, NC 27708, USA}
 \affiliation{School of Computational Science and Engineering,
 Georgia Institute of Technology, Atlanta, Georgia 30332, USA}
\affiliation{Department of Electrical and Computer Engineering, Duke University, Durham, NC 27708, USA}
\affiliation{Department of Chemistry, Duke University, Durham, NC 27708, USA}
\begin{abstract}
We simulate four quantum error correcting codes under error models inspired by realistic noise sources in near-term ion trap quantum computers: $T_2$ dephasing, gate overrotation, and crosstalk. We use this data to find preferred codes for given error parameters along with logical error biases and a pseudothreshold which compares the physical and logical gate failure rates for a CNOT gate. Using these results we conclude that Bacon-Shor-13 is the most promising near term candidate as long as the impact of crosstalk can be mitigated through other means.
\end{abstract}

\pacs{Valid PACS appear here}% PACS, the Physics and Astronomy
                             % Classification Scheme.
%\keywords{Suggested keywords}%Use showkeys class option if keyword
                              %display desired
\maketitle

\section{Introduction}
Quantum computing experiments have already demonstrated state stabilization \cite{fluhmann2019encoding, riste2015detecting, andersen2019entanglement,negnevitsky2018repeated},  single-axis quantum error correction~\cite{NMR3, chiaverini2004realization, SchindlerRepCpde2011,taminiau2014universal, rosenblum2018fault, kelly2015state}, multi-axis fault-tolerant quantum error detection~\cite{Linke422Ions2016, TakitaPRL2017, harper2019fault}, and we expect to be implementing full quantum error correction soon~\cite{wright2019benchmarking, trout2018simulating, OBrienSurface17DensityMatrix2017, BermudezITQCwithSteane2017}.  These small quantum devices will be the predecessors to far larger fault tolerant quantum computers which can run interesting algorithms at high rates of success~\cite{QEC1, QEC2, QEC3, litinski2018game}.

One of the first important uses of these new devices will be to better understand the actual errors they face~\cite{iyer2018small, harper2019efficient}. This information will be used to find optimally performing codes and decoders for the error model of a given architecture, leading to improved logical performance. With this goal in mind, we study the performance of four [[$9$,$1$,$3$]] quantum error-correcting codes under a set of error models that are common to ion trap quantum computing systems. The codes being considered here are the 17-qubit rotated surface code~\cite{TomitaLowDSC2014}, the 13-qubit Bacon-Shor code~\cite{BaconBaconShor2006, aliferis2007subsystem, li2018direct}, and two variants of Shor's code~\cite{QEC1}. There are many small QEC codes such as the [[$5$,$1$,$3$]] code~\cite{bennett1996mixed,Laflamme5qubit1996}, Steane [[$7,1,3$]] code~\cite{SteaneSteaneCode1996}, Bare [[$7,1,3$]] code~\cite{LiBareAnc2017}, twisted surface code~\cite{YoderSurfaceCodeTwist2017}, and tailored codes for biased error~\cite{robertson2017tailored} that can be implemented using 10-20 qubits, with pseudothresholds that have been improved by the introduction of flag qubits~\cite{chao2018quantum, reichardt2018fault, chamberland2018flag, lao2019fault}. Here we picked our set of codes to be gauge fixes within the 2-D quantum compass code model~\cite{li20192d}. As a result these codes require only bare ancilla for fault-tolerance and have high circuit-level pseudothresholds.

While general quantum error correction literature considers the depolarizing error model~\cite{KnillKnillEC2005,RaussendorfClusterState12007}, in reality errors emerging in quantum systems are expected to be more architecture dependent. Hence when studying the performance of algorithms and error correcting schemes in realistic systems we have to take into account the errors that are dominant in the given architecture. We consider an ion trap quantum computer that defines the qubit using hyperfine clock states of $^{171}$Yb$^{+}$~\cite{YbPrep}. In this case $T_1 > 10^{10}s$, so we can ignore its effects over the course of an experiment. Although single-qubit gates in similar systems have been shown to have fidelity beyond the error-correction pseudothreshold of these small QEC codes~\cite{NoekHiFidSPAMYb2013,CompPulseYb}, there are a couple of limiting factors for two-qubit operation fidelities and qubit lifetimes in general that are native to the ion trap system. The most common sources of error to consider in a trapped-ion system are $T_2$ dephasing errors, motional mode heating errors in the trap, and overrotation and crosstalk errors induced by the application of gates via lasers~\cite{wu2018noise}.

Of these sources of noise, $T_2$ and overrotation noise are both shared among most qubit implementations, however the actual model for crosstalk noise is very architecture specific. As progress has been made towards increasing qubit count and improving control, these unwanted qubit-qubit interactions known as crosstalk errors have become a significant error source in near-term quantum devices~\cite{merrill2014transformed}. In a trapped ion system, laser intensity spillover onto the neighboring ions during gate applications can lead to unwanted $XX$-type crosstalk errors between qubits involved in the desired gate and their neighbors in the ion chain. In a system implementing a small quantum error correcting code, such crosstalk errors can break fault tolerance and directly give rise to logical errors on the encoded information unless carefully dealt with. Therefore, in near-term quantum error correction experiments, steps must be taken to mitigate the damaging effects of crosstalk errors. Here we show that using a dynamic programming algorithm we can find optimal qubit to ion mappings for certain QEC codes that suppress the most damaging effects of first order crosstalk errors. 
\begin{figure*}
    \centering
    \includegraphics[height=0.218\linewidth]{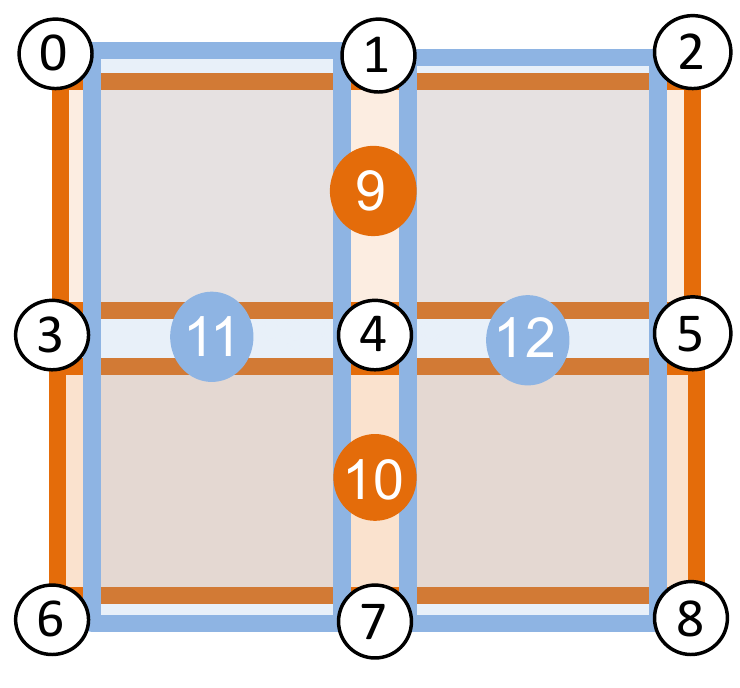}\hspace{1ex}
    \includegraphics[height=0.22\linewidth]{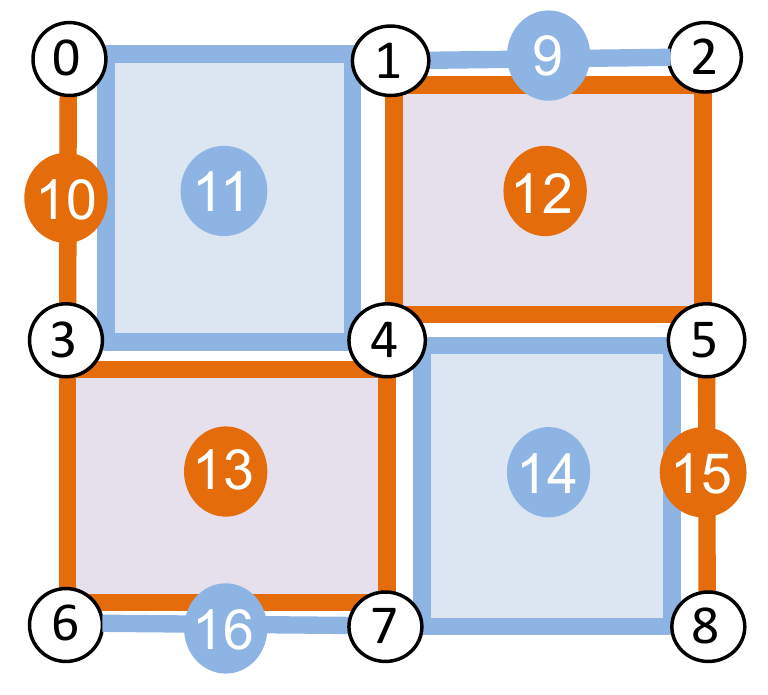}\hspace{1ex}
    \includegraphics[height=0.22\linewidth]{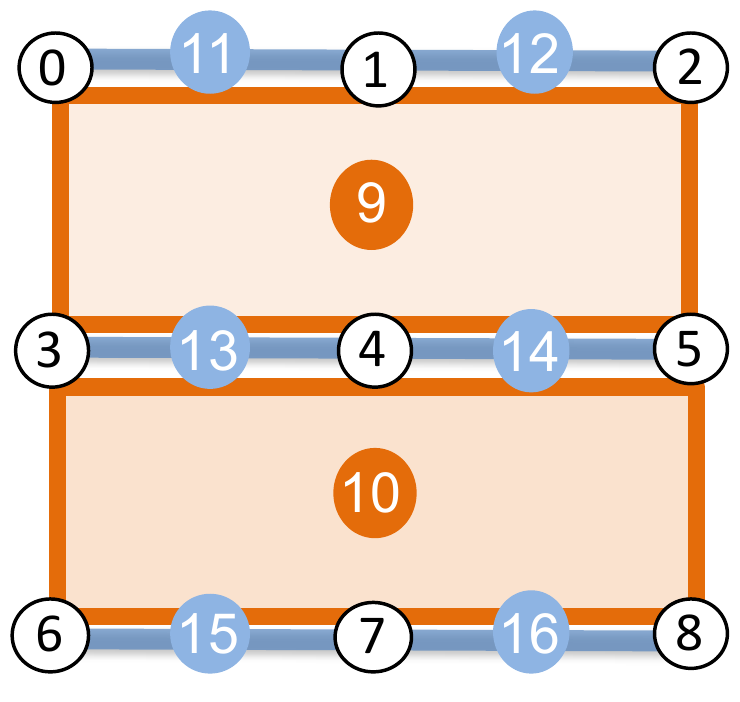}\hspace{1ex}
    \includegraphics[height=0.216\linewidth]{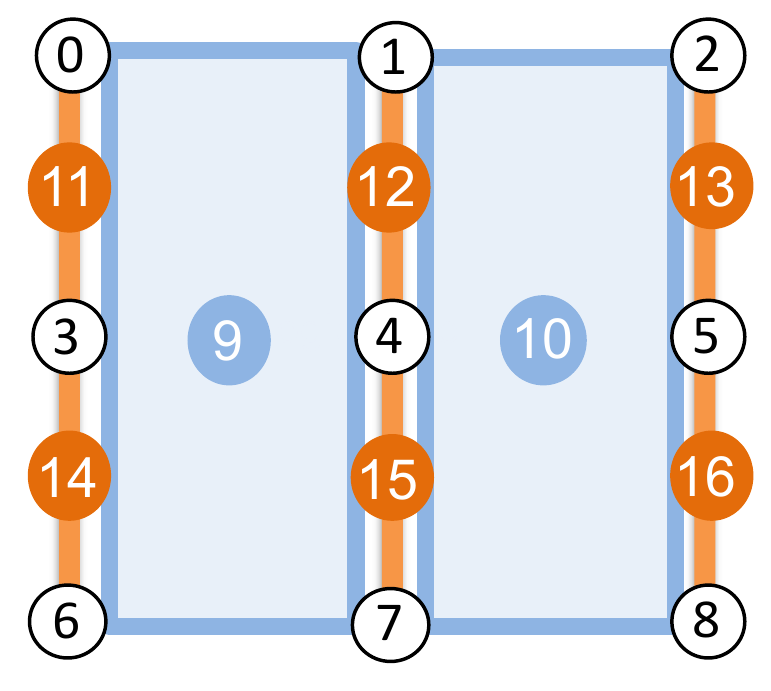}
    \caption{Stabilizer diagrams for (left to right) Bacon-Shor-13, Surface-17, Shor-6X2Z, and Shor-6Z2X. The orange connections/plaquettes represent $Z$-type stabilizers and the blue connections/plaquettes represent $X$-type stabilizers. White circles represent data qubits and colored circles represent ancillary qubits which measure the stabilizer they are attached to.}
    \label{fig:codes}
\end{figure*}
Previous experimental work using trapped ions has already demonstrated implementation of classical error correction~\cite{ChiaveriniBitFlipZZXXZX2004, SchindlerRepCpde2011}, fault-tolerant quantum error detection~\cite{Linke422Ions2016} and logical state encoding of quantum error correcting codes~\cite{NiggSteaneEnc2014, TakitaPRL2017}. Several theoretical studies have examined possibilities of implementing quantum error correcting codes in near-term experiments using trapped ions, including architectural studies of connecting multiple traps via ion shuttling or optical interconnects~\cite{WinelandExpIssueIons1998, KielpinskiQCCD2002, LekitscheMicroQCCD2017, MoehringEntangleIonQubitPhoton2007, DuanEntanglePhotonColloq2010, MonroeMUSIQC2014}, while others have looked at logical performances of small error correction codes in realistic error models \cite{trout2018simulating, li2018direct, brown2018comparing, BermudezITQCwithSteane2017, gutierrez2019transversality, bermudez2018fault}.

In this paper we study the logical performance of a transversal CNOT gate between two logical qubits that are maintained in the same trap under realistic error models featuring the mixing of overrotation, $T_2$ dephasing, and crosstalk. We present regions in which each code would perform the best in a near term experiment, along with the regions where the encoded qubit outperforms its physical counterpart. Our hope is that by finding optimal codes for these varied error models we will be able to hasten the arrival of successful error correcting implementations, which would be a major milestone in the pursuit of fault-tolerant quantum computing.

This paper is organized as follows: in Section~\ref{sec:codes} we briefly introduce the quantum error correcting codes that we consider, in Section~\ref{sec:noise} we explain the error models we use to simulate the noise being seen in the labratory setting, in Section~\ref{sec:crosstalk} we present the methods we use to mitigate the impact of ion trap crosstalk errors on encoded circuits, and in Section~\ref{sec:results} we show the numerical simulation results of code performances for different error parameters, along with logical error biases.
\section{Error Correction in a linear Ion Trap}\label{sec:codes}
\subsection{9 qubit compass codes} 
The quantum compass model on a square lattice of spins can be defined with the following Hamiltonian
\begin{equation*}
    H = \sum_i \sum_{j \neq L-1} J_X X_{i,j} X_{i,j+1} + \sum_{i \neq L-1} \sum_{j}J_Z Z_{i,j} Z_{i+1,j},
\end{equation*}
where $ZZ$ type interactions occur on spins linked by a vertical edge, and $XX$ type interactions for those sharing a horizontal edge~\cite{li20192d}. Quantum error-correcting codes can be defined using the method of gauge fixing: inserting sets of these two-qubit gauge operators into the stabilizer group by fixing the eigenvalue of their products. These codes have a number of nice features: all of their stabilizers can be measured fault-tolerantly using bare ancillas, they can be modified to deal with spatially asymmetric noise, and they are easily decoded. In this paper we consider 4 different $[[9,1,3]]$ quantum error-correcting codes that can be defined using the compass model defined by a $3 \times 3$ square lattice: the rotated 17-qubit surface code (Surface-17)~\cite{TomitaLowDSC2014}, the Bacon-Shor code (Bacon-Shor-13)~\cite{BaconBaconShor2006, LiBareAnc2017}, and two variations of Shor's code (Shor-6Z2X, Shor-6X2Z)~\cite{QEC1}. To clarify the code orientations we use, the stabilizers and logical operations of these codes are listed in Table~\ref{table:codes}. All of these codes can be implemented on a linear ion chain using at most 17 qubits to protect one logical qubit of quantum information.
\begin{table*}
\begin{center}
\begin{tabular}{ c|c|c|c}
 \hline
 Bacon-Shor-13 & Shor-6Z2X & Shor-6X2Z & Surface-17\\ \hline \hline
 \multicolumn{4}{c}{Stabilizers}\\ \hhline{----} \hline
$Z_0 Z_3 Z_1 Z_4 Z_2 Z_5$ & $X_0 X_1 X_3 X_4 X_6 X_7$ & $Z_0 Z_3 Z_1 Z_4 Z_2 Z_5$ & $Z_1 Z_2 Z_4 Z_5$\\ 
$Z_3 Z_6 Z_4 Z_7 Z_5 Z_8$ & $X_1 X_2 X_4 X_5 X_7 X_8$ & $Z_3 Z_6 Z_4 Z_7 Z_5 Z_8$ & $Z_0 Z_3$\\ 
$X_0 X_1 X_3 X_4 X_6 X_7$ & $Z_0 Z_3$ & $X_0 X_1$ & $Z_3 Z_4 Z_6 Z_7$\\ 
$X_1 X_2 X_4 X_5 X_7 X_8$ & $Z_1 Z_4$ & $X_1 X_2$ & $Z_5 Z_8$\\  
                        & $Z_2 Z_5$ & $X_3 X_4$ & $X_0 X_1 X_3 X_4$\\
                        & $Z_3 Z_6$ & $X_4 X_5$ & $X_6 X_7$\\  
                        & $Z_4 Z_7$ & $X_6 X_7$ & $X_4 X_5 X_7 X_8$\\  
                        & $Z_5 Z_8$ & $X_7 X_8$ & $X_1 X_2$\\ \hline \hline   
 \multicolumn{4}{c}{Logical Operators}\\ \hhline{----} \hline
 $ Z_0 Z_1 Z_2 $ & $Z_0 Z_1 Z_2$ &  $Z_0 Z_1 Z_2$ & $Z_0 Z_4 Z_8$ \\ 
 $ X_0 X_3 X_6 $ & $X_0 X_3 X_6 $ & $X_0 X_3 X_6 $ & $X_2 X_4 X_6$ \\ \hline
\end{tabular}
\caption{Stabilizers and logical operators of Bacon-Shor-13, Shor-6Z2X, and Shor-6X2Z, and Surface-17.}
\label{table:codes}
\end{center} 
\end{table*}
\subsection{Gate Implementations}
In this work we start with Clifford circuits composed of the gate set $\{X,H,CNOT\}$ along with preparation into the $\ket{0}$ state and measurement in the $Z$ basis. It should be noted that since we do not allow preparation into $\ket{+}$, certain logical states require more single qubit gates than others to be prepared, leading to worse error rates. Once we have these circuits, we decompose them into ion-trap gates using the identities in Figure~\ref{fig:gates}:
\begin{figure}[h!]
    \centering
    \resizebox{.65\linewidth}{!}{
    \mbox{
    \Qcircuit @C=.9em @R=.7em {
     & \gate{H} & \qw & \push{\rule{.3em}{0em}=\rule{.3em}{0em}} & & \gate{RX(-\pi)} & \gate{RY(\frac{\pi}{2})} & \qw\\
    & & & \push{\rule{.3em}{0em}=\rule{.3em}{0em}} & & \gate{RY(-\frac{\pi}{2})} & \gate{RX(\pi)} & \qw
    }
    }}
    \vspace{1.5em}
    
    \resizebox{\linewidth}{!}{
    \Qcircuit @C=.9em @R=.7em {
     & \ctrl{1} & \qw & \raisebox{-2.5em}{=} & & \gate{RY(v\frac{\pi}{2})} & \multigate{1}{XX(s\frac{\pi}{4})} & \gate{RX(-s\frac{\pi}{2})} & \gate{RY(-v\frac{\pi}{2})} & \qw\\
   & \targ & \qw & & & \qw & \ghost{XX(s\frac{\pi}{4})} & \gate{RX(-vs\frac{\pi}{2})} & \qw & \qw
    }}
    \caption{Ion trap gate compilations of CNOT and $H$ in terms of one- and two-qubit Pauli rotations~\cite{MaslovCircuitCompIT2017}. The choices of $s,v \in \{\pm 1\}$ represent degrees of freedom that only affect the global phase.}
    \label{fig:gates}
\end{figure}
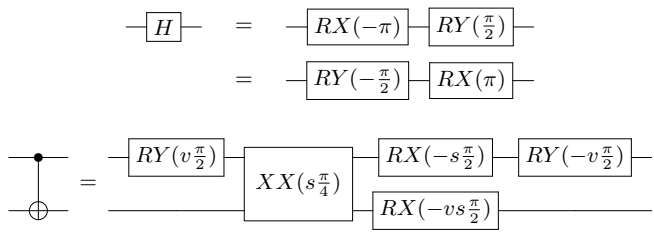
where  $RX(\theta)=exp(-i\frac{\theta}{2}X)$, $RY(\theta)=exp(-i\frac{\theta}{2}Y)$, and  $XX(\theta)=exp(-i\theta XX)$. $s,v = \pm 1$ and we choose the sign to cancel as many single qubit gates as possible~\cite{MaslovCircuitCompIT2017}. We think of our gates being applied through a Rabi frequency which is evolved for a time such that $\theta = \Omega_{R}t$.   One of the benefits of the ion-trap architecture is that the two qubit M\o{}lmer-S\o{}rensen gates, $XX(\pi/4)$~\cite{MolmerPRL1999, SorensenPRLMSgate1999} can be applied between distant qubits in the chain~\cite{leung2018entangling, LeungFMgates2017, zhu2006arbitrary, blumel2019power, landsman2019two}. This allows us to not only avoid the SWAP gates that other architectures rely on, but also gives us the freedom to label our ions as we desire. We will use this freedom reduce the impact of crosstalk in Section~\ref{sec:crosstalk}. It is on these ion trap circuits that we apply our noise models.
\section{Noise models} \label{sec:noise}
The most general error model we consider is the depolarizing model, which applies random Pauli errors after gates:
\begin{align*}
\begin{split}
E_{1d} &= \{\sqrt{1-p_{1d}}I, \sqrt{p_{1d}/3}X, \sqrt{p_{1d}/3}Y, \sqrt{p_{1d}/3}Z\}\\
E_{2d} &= \{\sqrt{1-p_{2d}}I, \sqrt{p_{2d}/15}IX, \ldots \sqrt{p_{2d}/15}ZZ\}
\end{split}
\end{align*}
Our model holds the single qubit error rate as one-tenth of the two qubit error rate:
$$P_{1d} = \frac{1}{10}P_{2d}.$$
This model has been studied many times and we mention it to provide a frame of reference when considering our results in relation to other work in the field.

The three ion trap specific error types we consider are $T_2$ dephasing, overrotation, and crosstalk. The first of these errors is an idling error, and the remaining two are gate errors. For all models we only consider stochastic channels due to limitations on memory within simulations for a logical two-qubit gate. 

\subsection{$T_2$ Dephasing}
We only consider idling error in the form of $T_2$ dephasing due to the long $T_1$ times in trapped ion systems. In this paper we look at $T_2$ times in the range of $$0 \leq \frac{1}{T_2} \leq 2 \: s^{-1}$$ for gate times of $10 \mu s$ and $200 \mu s$ for single and two qubit gates respectively. We also will allow for parallel single qubit gates, but only allow one two-qubit gate to be active at a time. This restriction is pessimistic given recent implementations of parallel gates~\cite{figgatt2019parallel}.  During the application of all one- and two-qubit gates we model a single-qubit dephasing error on each of the idling qubits (qubits that are not affected by the gate in operation) with the Kraus channel
\begin{subequations}
\begin{equation}\label{eq:t2Kraus}
    \begin{split}
        E_{idle} = \{ \sqrt{1-p_i}I, \sqrt{p_i}Z\},
    \end{split}
\end{equation}
where
\begin{equation}\label{eq:crosstalkProb}
    p_i = \frac{1}{2}\left( 1 - \exp\left[-\frac{1}{2}\frac{T_{idle}}{T_2}\right]\right),
\end{equation}
\end{subequations}
and $T_{idle}$ is the idling time of the particular qubit.
\subsection{Gate Error}
The next form of error we will consider are gate errors inspired by overrotation. These errors occur on any one or two qubit gate applied in ion traps, are one of our dominant sources of error~\cite{trout2018simulating}, and can stem from sources such as incorrect timing or miscalibrated laser intensities that lead to fluctuations in Rabi frequency. For single qubit gates, we can use composite pulse sequences to suppress the error~\cite{brown2004arbitrarily}, but for two qubit gates these sequences take prohibitive amounts of time. In certain cases these multi-qubit gate errors can be dealt with effectively as we will discuss in the next section, but in general we will have to rely on error correction to fix these errors.
In the overrotation error model, the gate error following some Pauli rotation gate $G$ has the form,\vspace{1ex}
\begin{equation}
    \varepsilon_G(\rho) = \kappa\cdot\varepsilon_G^c(\rho) + (1 - \kappa)\cdot\varepsilon_G^s(\rho)
    \label{errchan}
\end{equation}\vspace{.1ex}
where $\varepsilon_G^c$ and $\varepsilon_G^s$ are coherent and stochastic overrotation channels with equal fidelity given by,\vspace{1ex}
\begin{equation}
    \begin{split}
        \varepsilon_G^c(\rho) &= \exp (-i\epsilon G)\rho\exp (i\epsilon G)\\
        \varepsilon_G^s(\rho) &= \cos^2(\epsilon)I\rho I + \sin^2(\epsilon)G\rho G.
    \end{split}
    \label{channels}
\end{equation}\vspace{.1ex} 
In this paper we focus on stochastic gate error channels where $\kappa = 0$. We model the stochastic overrotation channels in the Clifford simulation as
\begin{align*}
    E_{MS} &= \{\sqrt{1-p_{MS}}I, \sqrt{p_{MS}}XX\}, \,\, p_{MS} = \sin^2(\epsilon_{MS})
\end{align*}
after M\o{}lmer-S\o{}rensen gate, and 
\begin{align*}
    E_{1q} &= \{\sqrt{1-p_{1q}}I, \sqrt{p_{1q}}P\}, \,\, p_{1q} = \sin^2(\epsilon_{1q})
\end{align*}
after single-qubit rotation gates $G \in \{R_X,R_Y,R_Z\}$.
% a single qubit Pauli $X$ would be followed by a gate
% $$\varepsilon_RX(\rho) = (1 - p_{1q})I\rho I + p_{1q}X\rho X$$
% and a two qubit M\o{}lmer-S\o{}rensen gate is followed by 
% $$\varepsilon_{XX}(\rho) = (1 - p_{2q})I\rho I + p_{2q}X_1X_2\rho X_1X_2.$$
This error model is less damaging in general than the coherent case, but these errors must be mitigated through error correction instead of creative compiling.

The challenge of correcting coherent overrotation errors in ion traps is interesting because the errors are in fact invertible. If one is able to apply the correct channel to the data, the error can have its damage undone. This is in contrast with stochastic error channels which cannot be inverted and as a result require projective measurement and correction in order to be dealt with. The technique of \textit{stabilizer slicing} handles coherent overrotations by taking advantage of the underlying stabilizer state nature of our logical codestates in order to direct overrotations against each other~\cite{debroy2018slicing}. In this way we can eliminate the impact of the errors stemming from stabilizer measurement before they can even be seen. Due to their symmetries being easily broken down into weight-$2$ operators the Shor codes are best suited for implementing stabilizer slicing using present day physical gates. The Bacon-Shor code can also implement slicing, but at slightly lower effectiveness over multiple rounds as the gauge wanders in time. Ref.~\cite{debroy2018slicing} shows that with increasing coherence in the error stabilizer slicing yields an improvement in single logical qubit error correction circuits for Bacon-Shor-13 and constant performance with coherence for Surface-17.

\subsection{Crosstalk} \label{sec:noiseCrosstalk}
\begin{figure}
    \centering
    \includegraphics[trim = 325 50 325 40, width = 0.9\linewidth]{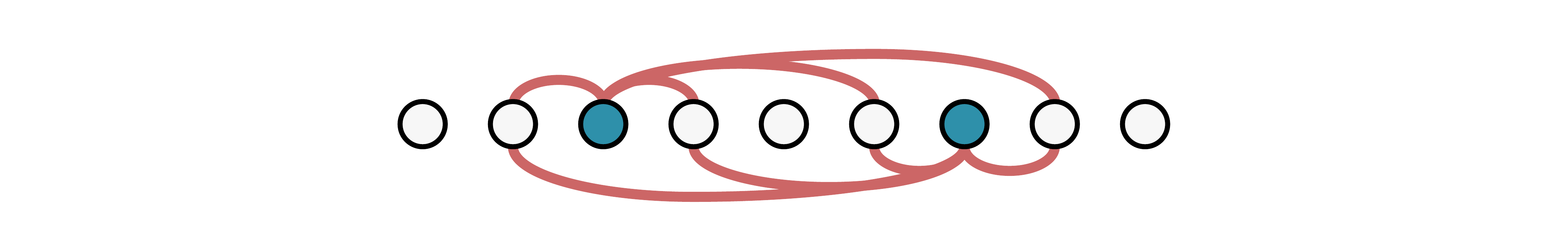}
    \caption{The first order crosstalk errors, shown in red, which occur during a M\o{}lmer-S\o{}rensen gate on the qubits shaded in blue.}
    \label{fig:crosstalk ions}
\end{figure}

\begin{figure}[h!]
    \centering
    \includegraphics[width = \linewidth]{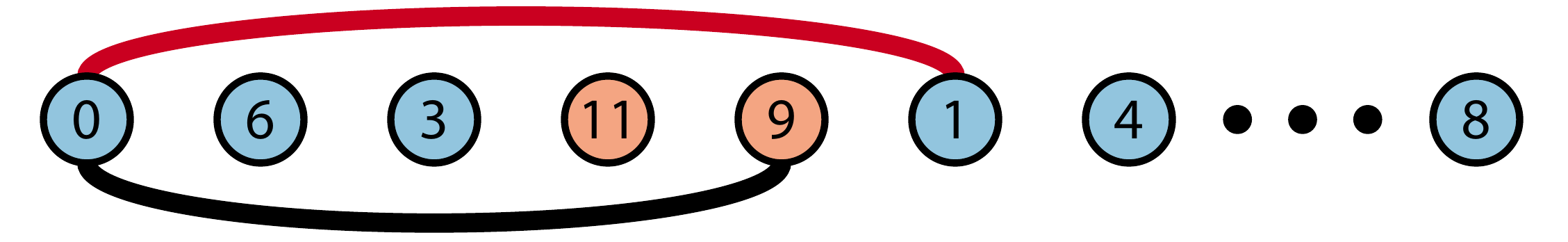}
    \caption{In this figure we show a possible chain to encode Bacon-Shor-13, where the data qubits are blue and the ancillae are red. When the M\o{}lmer-S\o{}rensen gate (denoted in black) is applied between qubits 0 and 9 as part of the $Z_0Z_3Z_1Z_4Z_2Z_5$ stabilizer, there is a first order crosstalk event which causes an $XX$ type error (in red) between qubits 0 and 1. This error will cause a logical error as Bacon-Shor-13 cannot differentiate between it and the weight-$1$ $X$-type error on qubit 2, and when this correction is implemented we would have applied a full $X$-type logical error.}
    \label{fig:BS Crosstalk}
\end{figure}

Finally, crosstalk is an issue that leads to pairwise correlated errors when applying our native entangling gate, the M\o{}lmer-S\o{}rensen gate. When an entangling gate is applied, a global beam is applied to the chain, and individually addressed beams are applied to the involved qubits. These addressed beams can have some degree of overlap with the neighboring qubits. For single qubit gates, this can easily be handled by narrowband or passband composite pulses~\cite{merrill2014transformed}.  For two-qubit gates, this leads to a possibility for small M\o{}lmer-S\o{}rensen type errors between the involved qubits and any of these nearest neighbors. We model this effect through applying a Kraus channel to all qubit pairs $\{q_i,q_n\}$ where $q_i$ is a qubit involved in the desired M\o{}lmer-S\o{}rensen gate, and $q_n$ is a qubit that neighbors either of the involved qubits in the physical ion chain. These pairs are shown in Figure~\ref{fig:crosstalk ions}. For each of these pairs the following Kraus channel is applied:
\begin{subequations}
\begin{equation}\label{eq:crosstalkKraus}
    \begin{split}
        E_{crosstalk} = \{ \sqrt{1-p_c}II, \sqrt{p_c}XX\},
    \end{split}
\end{equation}
where
\begin{equation}\label{eq:crosstalkProb}
    p_c = \sin^2(\frac{\Omega_c}{\Omega_R} \times \frac{\pi}{4}).
\end{equation}
\end{subequations}
$\Omega_c /\Omega_R$ is the two-qubit gate crosstalk Rabi ratio, which gives the ratio of the Rabi frequency experienced by these crosstalk pairs and the Rabi frequency of the intended gate. Under this model, a single M\o{}lmer-S\o{}rensen gate can lead to 8 possible first order M\o{}lmer-S\o{}rensen type crosstalk errors when the qubits are well separated. If the intended gate is being applied on two qubits with only a single qubit separating them, the effect increases dramatically and crosstalk errors featuring this central qubit occur at four times their usual rate.
These large scale correlated errors can cause issues with fault tolerance, and chain orderings which do not account for them may have possible first order crosstalk events that lead to a logical error, as shown in Figure~\ref{fig:BS Crosstalk}. We explain methods for avoiding these damaging crosstalk events in the following section. In this paper we consider stochastic crosstalk, however in the case of coherent crosstalk one can use Pauli conjugation to control the impact of these errors~\cite{cai2019mitigating}, along with dynamical decoupling methods~\cite{viola1999dynamical}.
\section{Fault-tolerance to crosstalk in ion chains} \label{sec:crosstalk}
In practice when we apply a two-qubit operation between the information stored in the $i$th and $j$th ion on an ion chain, the four neighboring ions $i-1, i+1, j-1$ and $j+1$ may be affected by the laser beam, which would introduce undesired crosstalk. In our stochastic model, a full $XX$ error can happen between the qubits stored on the following pairs of ions $(i-1, i)$, $(i+1, i)$,$(j-1, i)$, $(j+1,i)$, $(i-1, j)$, $(i+1, j)$, $(j-1, j)$, $(j+1, j)$, which we will refer to as the crosstalk pairs. The $XX$ error happens on each pair with probability $p_c$ defined in Eq.~\ref{eq:crosstalkProb} and pairs for which crosstalk can induce a logical error will be referred to as bad crosstalk pairs.
    
There are multiple classes of bad crosstalk errors. Crosstalk errors that impact two data qubits, as in Figure~\ref{fig:BS Crosstalk}, can clearly cause a distance drop. Additionally, crosstalk errors which apply to both a data qubit and an ancillary qubit such that the $X$ error propagates back to a data qubit can also cause issues. The last case is one in which two ancillary qubits have a crosstalk error between them which causes $X$ errors to propagate to the data, causing a logical error. This type of error is avoided in our circuit by having all of our stabilizer measurements serialized, so the correlated errors do not propagate in dangerous ways. Our particular circuit compilation is also set up so that crosstalk errors are never conjugated into weight-2 $Z$-type errors, so $Z$-type logical errors are not as much of a concern in our crosstalk pairs.

\begin{figure}
    \centering
    \includegraphics[width = 0.48\linewidth]{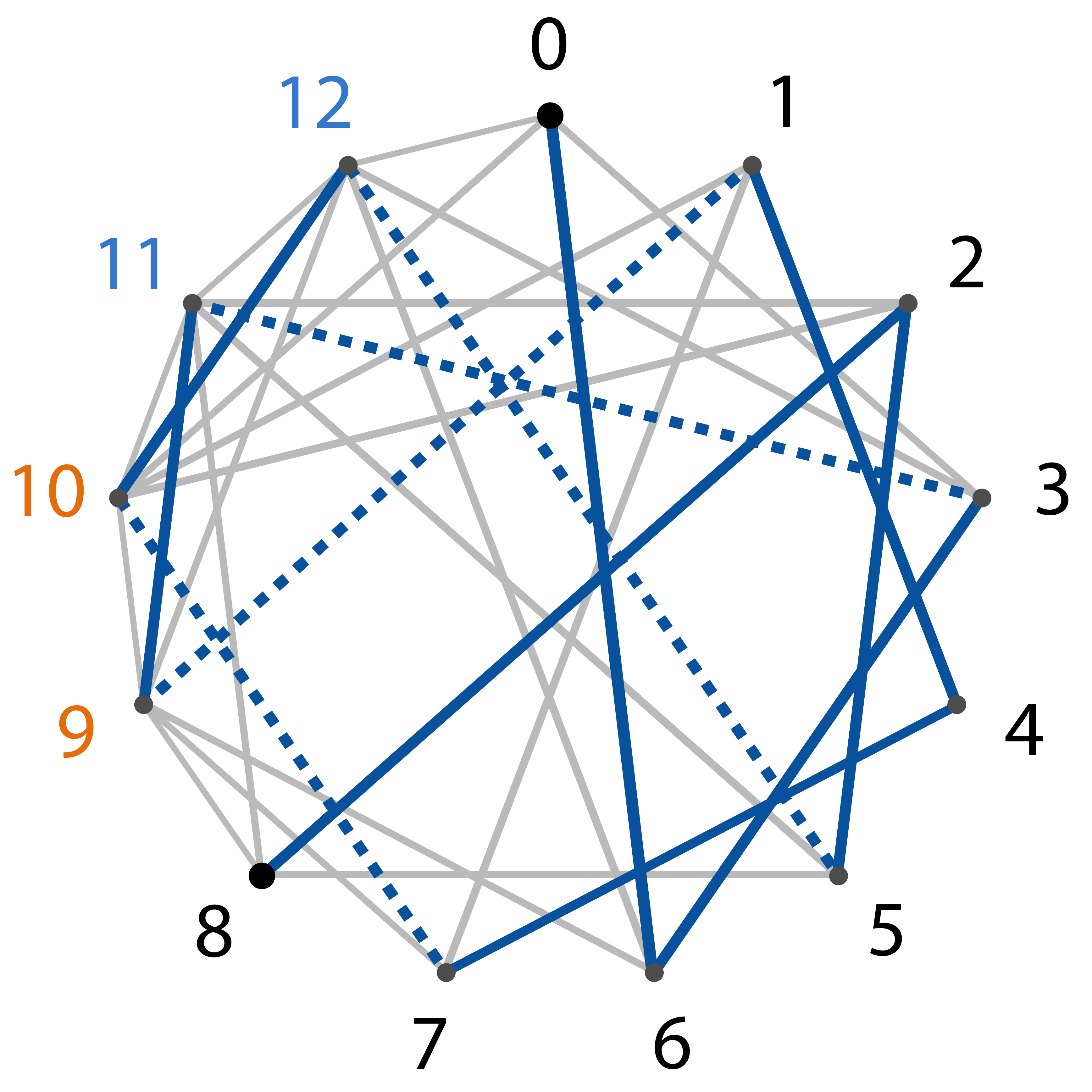}\hspace{2ex}\includegraphics[width = 0.48\linewidth]{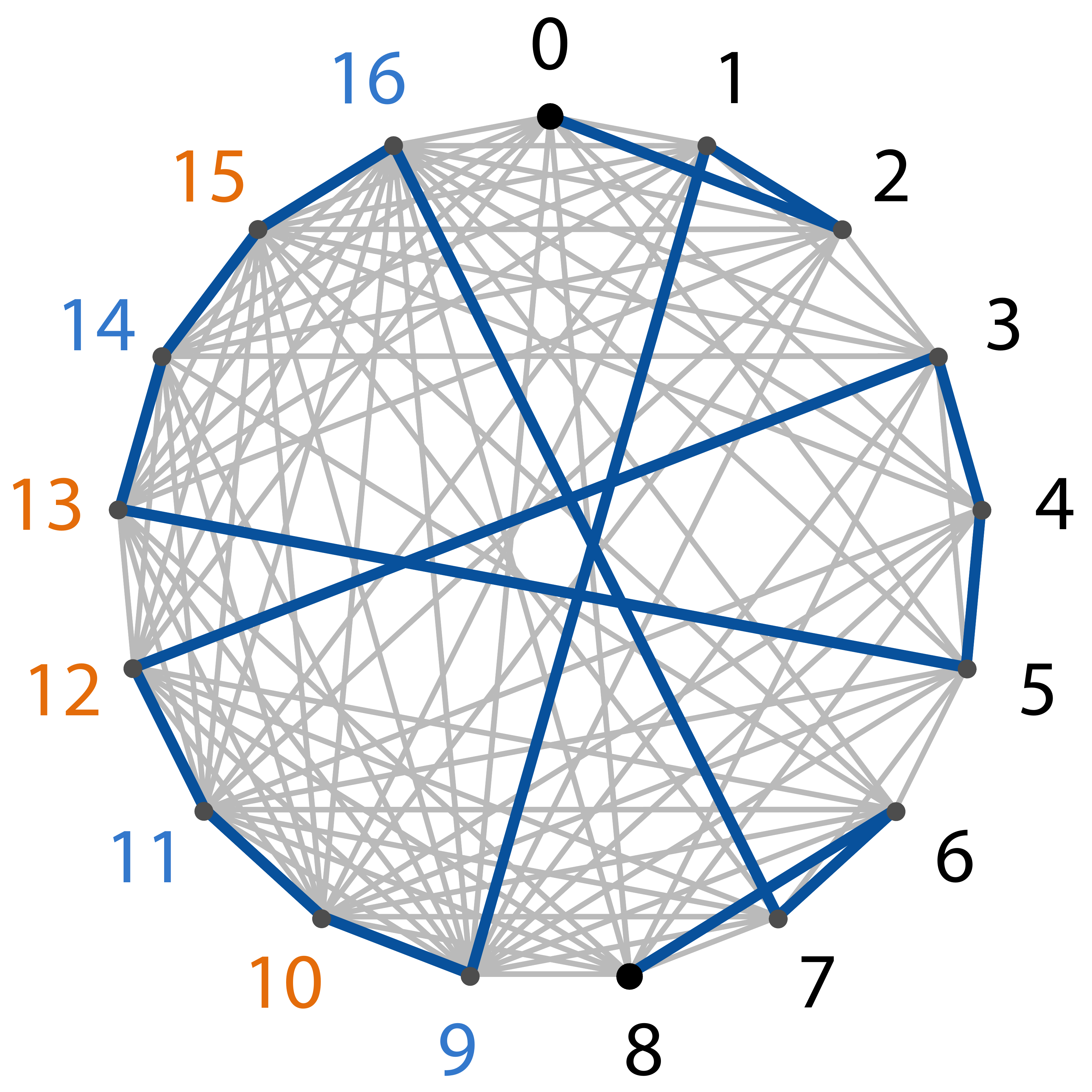}
    \caption{Graphs for Bacon-Shor-13 (left) and Surface-17 (right) where the edges correspond to qubits which could be neighbors without leading to distance-damaging crosstalk errors. We use dynamic programming to find a Hamiltonian path for each graph (blue lines) which also minimizes operation time. A Hamiltonian path is not possible for Bacon-Shor-13, so we must add in extra connections (dotted lines) which introduce distance-damaging crosstalk errors. Data qubits are labeled in black and ancillary qubits are labeled blue(orange) to indicate they measure $X$($Z$)- type stabilizers.}
    \label{fig:crosstalk paths}
\end{figure}

In order to be robust against these crosstalk errors when implementing a small quantum error-correcting code on an ion trap quantum computer, we try to find an optimal mapping of qubits on the linear ion chain such that a single $XX$ error event on any of the possible crosstalk pairs does not directly lead to a failure on the encoded logical state. Our problem can be formulated using the graph theory language: we construct a graph $G = (V,E)$, where the vertex set $V$ is the set of qubits, and the edge set $E$ consists of pairs of qubits such that mapping these two qubits as neighbors will not lead to any bad crosstalk pairs. One can find an ion chain without any bad crosstalk pairs by first finding a path 
$$q_{i_1} \rightarrow q_{i_2} \rightarrow \cdots q_{i_j} \rightarrow \cdots \rightarrow q_{i_{|V|}}$$
in the graph that covers each vertex (qubit) exactly once, then mapping the qubit $q_{i_j}$ to the $j$th ion of the chain. In Figure~\ref{fig:crosstalk paths} we show the graphs and corresponding path solutions for Bacon-Shor-13 and Surface-17. Note that the ancillary qubits have more connections, as the way we compile our circuits means that data-data crosstalk is the most damaging effect.

The problem of traversing a graph and crossing each vertex once is known as the Hamiltonian Path problem~\cite{book1980michael}. Although Hamiltonian Path is NP-complete, we can use techniques such as dynamic programming to accelerate brute-force searching~\cite{bellman1961dynamic}. See Appendix~\ref{sec:appendixDP} for further details.
\subsection{Best Chains for Different Codes}
Since the structure of stabilizer measurements is code-specific, the constraints that the proper ion chain needs to satisfy are also different. Consequently each chain ends up being different. Also note that all the chains presented in this section are also designed to minimize execution time for the corresponding circuit as a secondary constraint. Data qubits are depicted in black and ancillary qubits are labeled blue(orange) to indicate they measure $X$($Z$)- type stabilizers.
\begin{itemize}
    \item \textbf{Surface-17}
        \begin{equation*}
            \textbf{0}\,\, \textbf{2}\,\, \textbf{1}\,\, {\color{xblue}\textbf{9}}\,\, {\color{zorange}\textbf{10}}\,\, {\color{xblue}\textbf{11}}\,\, {\color{zorange}\textbf{12}}\,\, \textbf{3}\,\, \textbf{4}\,\, \textbf{5}\,\, {\color{zorange}\textbf{13}}\,\, {\color{xblue}\textbf{14}}\,\, {\color{zorange}\textbf{15}}\,\, {\color{xblue}\textbf{16}}\,\, \textbf{7}\,\, \textbf{6}\,\, \textbf{8}
        \end{equation*}
        
    \item \textbf{Bacon-Shor-13}
        \begin{equation*}
            \textbf{0}\,\, \textbf{6}\,\, \textbf{3}\,\, {\color{xblue}\textbf{11}}\,\, {\color{zorange}\textbf{9}}\,\, \textbf{1}\,\, \textbf{4}\,\, \textbf{7}\,\, {\color{zorange}\textbf{10}}\,\, {\color{xblue}\textbf{12}}\,\, \textbf{5}\,\, \textbf{2}\,\, \textbf{8}
        \end{equation*}
        
    \item \textbf{Shor-6X2Z}
        \begin{equation*}
            \textbf{0}\,\, \textbf{2}\,\, \textbf{1}\,\, {\color{xblue}\textbf{11}}\,\, {\color{xblue}\textbf{12}}\,\, {\color{zorange}\textbf{9}}\,\, {\color{xblue}\textbf{13}}\,\, \textbf{3}\,\, \textbf{4}\,\, \textbf{5}\,\, {\color{xblue}\textbf{14}}\,\, {\color{zorange}\textbf{10}}\,\, {\color{xblue}\textbf{15}}\,\, \textbf{6}\,\, \textbf{7}\,\, \textbf{8}\,\, {\color{xblue}\textbf{16}}
        \end{equation*}
        
    \item \textbf{Shor-6Z2X}
        \begin{equation*}
            \textbf{3}\,\, {\color{zorange}\textbf{11}}\,\, \textbf{0}\,\, \textbf{6}\,\, {\color{zorange}\textbf{12}}\,\, \textbf{1}\,\, \textbf{7}\,\, {\color{zorange}\textbf{13}}\,\, {\color{xblue}\textbf{9}}\,\, {\color{xblue}\textbf{10}}\,\, {\color{zorange}\textbf{14}}\,\, \textbf{4}\,\, {\color{zorange}\textbf{15}}\,\, \textbf{2}\,\, \textbf{8}\,\, {\color{zorange}\textbf{16}}\,\, \textbf{5}
        \end{equation*}
\end{itemize}
For Bacon-Shor-13 and the Shor codes unfortunately there do not exist any ion chain arrangement that could avoid all logical crosstalk errors, so the above chains are the ones which minimizes the impact of crosstalk errors to the system.

These codes do not have valid distance-preserving chains because they feature large weight-$6$ stabilizers. This means the two qubits neighboring the ancilla for these stabilizers must be acceptable crosstalk pairs with a large number of other qubits, and within the solution space provided by these small codes, there simply is not enough freedom to find a valid ordering. An alternative approach is to add additional spacer ions that could also be used for sympathetic cooling~\cite{KielpinskiSympTheory2000,BarrettSymCoolLogic2003,home2009memory,wang2017single}
\section{Results and Discussion} \label{sec:results}
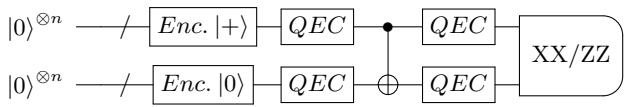
\begin{figure}[h]
    \centering
    \mbox{
    \Qcircuit @C=1em @R=.8em {
            \lstick{\ket{0}^{\otimes n}} & \qw & {/} \qw & \gate{Enc.\: |+\rangle} & \gate{QEC} & \ctrl{1} & \gate{QEC} & \multimeasureD{1}{\text{XX/ZZ}}\\
            \lstick{\ket{0}^{\otimes n}} & \qw & {/} \qw & \gate{Enc.\: |0\rangle} & \gate{QEC} & \targ & \gate{QEC} & \ghost{\text{XX/ZZ}}
            }
    }
    \caption{The circuit that we simulate for each code. From the $XX$($ZZ$) measurements we can gauge the code's performance in generating our desired state of $\Phi^+_L = \frac{1}{2}(|00\rangle_L + |11\rangle_L)$.}
    \label{fig:circuit}
\end{figure}
In order to assess the performance of a code against a given error model, we use a stabilizer method with importance sampled error~\cite{LiBareAnc2017} to simulate a circuit featuring both an $X$ and $Z$ basis state preparation, along with the ex-Rec of a transversal CNOT, as shown in Figure~\ref{fig:circuit}. By measuring the $XX$ and $ZZ$ parities of the output state, we can assess the performance of the code in a way which is experimentally implementable while also including only fault tolerant circuits. Logical $Y$ cannot be measured fault tolerantly since we are unable to do a round of classical correction after measuring the data qubits in the $Y$ basis. One thing to note is that the Shor's code variants do not possess a fault-tolerant $H$. As a result, in Bacon-Shor-13 and Surface-17 both bases are similarly difficult to prepare, while in the Shor's code variants one basis has an encoding circuit and the other requires projective preparation. Despite this lack of a fault-tolerant $H$ gate, since each Shor's code variant is CSS we can measure them in both the $X$ and $Z$ bases. We use these measurements to define a circuit level error version of pseudothreshold where the logical performance on this circuit is compared to the error rate for the unencoded CNOT:
\begin{equation}
        p_{phys} = p_{2q} + 4 \cdot p_{1q} + 8\cdot p_{crosstalk} + 2\cdot p_{idle}
    \label{eq:phys err}
\end{equation}
where
\begin{equation}
    \begin{split}
        p_{idle} &= \frac{1}{2}\left( 1 - \exp\left[ -\frac{t_{2q} + 3t_{1q}}{2T_2}\right]\right)\\
        p_{crosstalk} &= \sin^2\left( \frac{\Omega_c}{\Omega_R} \times \frac{\pi}{4}\right)
    \end{split}
\end{equation}
and $p_{2q} = 10\:p_{1q}$, $T_2$, and $\Omega_c /\Omega_R$ are the three error parameters we will be studying. We include a factor of two on $p_{idle}$ because there are at least two qubits experiencing the idling, and the factor of eight on the crosstalk probability comes from the idea that there are eight possible first order crosstalk pairs, as shown in Figure~\ref{fig:crosstalk ions}. Due to this non-standard definition of pseudothreshold, certain pseudothresholds will seem very low because the comparison we are making includes the impact of preparation circuits and transversal gates.
\subsection{Depolarizing Error Model} \label{sec:depolarizing}
\begin{figure}[h!]
    \centering
    \includegraphics[trim = 50 0 15 0, width = 0.85\linewidth]{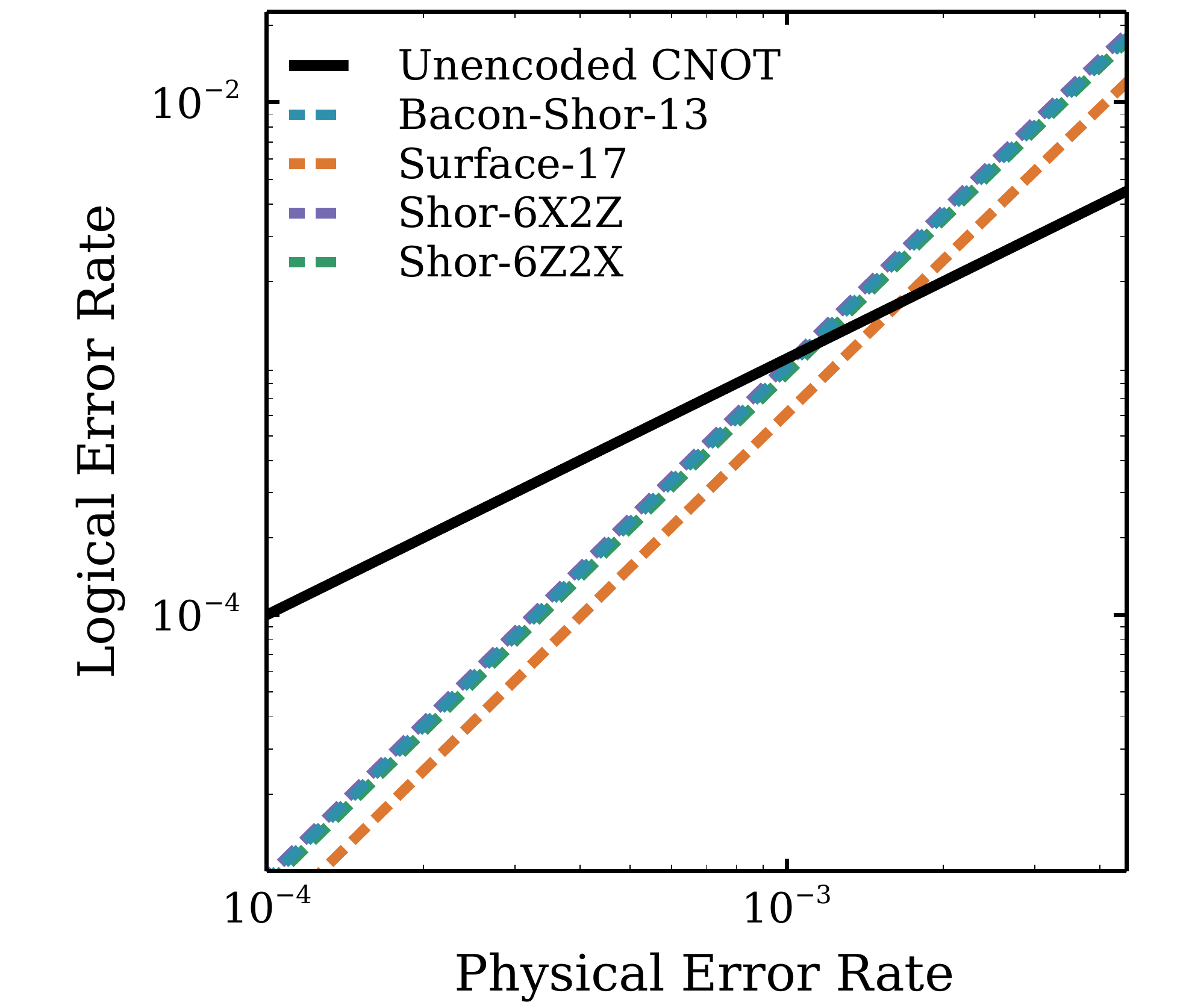}
    \caption{In this plot we compare the physical two qubit error rate to the error rate on the circuit in Figure~\ref{fig:circuit} under the standard depolarizing model described in Section~\ref{sec:noise}. }
    \label{fig:TCNOT Standard}
\end{figure}
In Figure~\ref{fig:TCNOT Standard} we consider the gate depolarizing error model defined in the beginning of Section~\ref{sec:noise}, with depolarizing errors directly following all gates but not acting on idles. Surface-17 outperforms all other codes due to its ability to correct both types of error. Bacon-Shor-13 does not have the same ability to correct two qubit errors, and both Shor's code variants share this weakness in one direction. These codes are more effective than Bacon-Shor-13 at correcting one side of error, however their projective preparation requirement for one of the logical bases cancels out this benefit. Shor-6Z2X outperforms its counterpart because the 6X2Z form of the circuit uses 64 additional single qubit gates, leading to extra error locations.
\subsection{Ion Trap Error Models} \label{sec:iontrap}
We now consider the remaining error models mentioned in Section~\ref{sec:noise}. In Figure~\ref{fig:Performance table} we present a series of phase diagrams indicating the transition between regions in which different codes are optimal choices. For the M\o{}lmer-S\o{}rensen vs. idling plots (a),(d) there are no crosstalk errors, and in the M\o{}lmer-S\o{}rensen vs. Rabi ratio plots (b),(e) we have set $T_2 = \infty$. However in the idling vs. Rabi ratio plots (c),(f) we have set $p_{2q} = 10\: p_{1q} = 0.0001$ since we believe it is unrealistic for there to be no error on the qubits involved in a gate.
\subsubsection{Code Performance}
\begin{figure*}[t!]
    \centering
    
    \includegraphics[width=\linewidth]{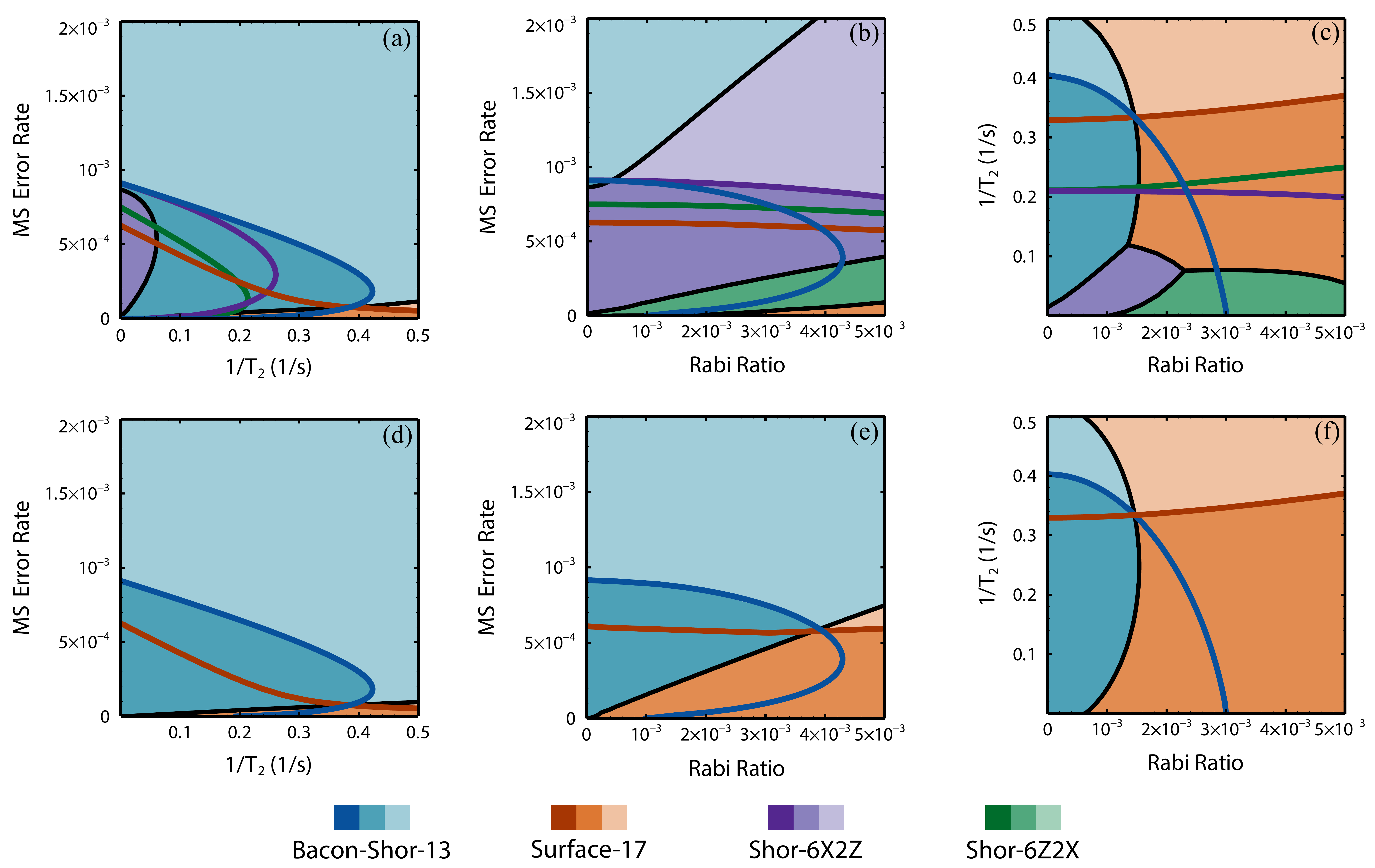}
    \caption{Figures showing best performing codes and pseudothresholds for different error models and sets of codes. In (a,b,c) we are comparing codes in the set \{Surface-17, Bacon-Shor-13, Shor-6X2Z, Shor-6Z2X\}, while in (d,e,f) we restrict our set to only consider codes with a transversal Hadamard gate, \{Surface-17, Bacon-Shor-13\}. In (a,d) we look at the intersection of overrotation error (parameterized by the two qubit gate error) and $T_2$ dephasing, in (b,e) we look at overrotation and crosstalk, and in (c,f) we look at $T_2$ dephasing and crosstalk with a background overrotation characterized by a M\o{}lmer-S\o{}rensen error rate of $10^{-4}$. The colored regions indicate which code is optimal at those error parameters, with darker shading implying the code is outperforming a physical CNOT. The colored curves are the pseudothreshold curves for which the logical error rate is equal to the physical error rate in Eq.~\ref{eq:phys err} and the black curves are borders between regions in which different codes are preferred.}
    \label{fig:Performance table}
\end{figure*}
In Figure~\ref{fig:Performance table} we present plots for a variety of error models showing optimal codes and circuit-level pseudothreshholds. There are a few key features that distinguish the codes.

First, Surface-17 performs very poorly under gate errors, and only begins to outperform Bacon-Shor-13 and the two Shor's code variants when the other two error models are strong relative to overrotation. The surface code's advantage on the other two error models is more pronounced in the case of crosstalk. This is due to the surface code allowing for a fault-tolerant chain ordering. For the $T_2$ times we consider it is only relevant when overrotation errors are practically nonexistent. As would be expected from this trend, the plot in which it performs best is the idling versus crosstalk plot, in which overrotation error is minimal.

Second, crosstalk seems to be the only situation in which Bacon-Shor-13 does not excel in our comparison. Some of this is from the fact that Bacon-Shor-13 has a fault-tolerant preparation circuit for both bases, while both the Shor's code variants need projective preparation for one basis and Surface-17 needs it for both. Bacon-Shor-13 is also good at handling overrotations as they become more coherent, so for our ion-trap error models we believe it to be the best choice as long as we can control crosstalk via other means.

Finally, Figure~\ref{fig:Performance table}b the data shows that Shor-6X2Z is better at dealing with crosstalk than Shor-6Z2X. This seems unintuitive as Shor-6Z2X is optimized for catching the $X$-type errors that crosstalk is slightly biased towards, however due to its stabilizer structure, Shor-6X2Z is able to completely ignore a number of weight-$2$ $X$ errors on data qubits due to them being in its stabilizer group. This effect leads to Shor-6X2Z being better at correcting pairwise correlated $X$-type errors even though Shor-6Z2X is preferable for single data qubit errors.

Other than best performing codes, these plots also include pseudothresholds. Due to our particular metric based on Figure~\ref{fig:circuit}, they look different than would be expected based on other work in this area. This discrepancy is especially noticeable in the case of $T_2$ dephasing, where it can be seen that as other error sources approach zero, the pseudothreshold decreases significantly. Most definitions of pseudothreshold with respect to $T_2$ dephasing compare the dephasing timescale of the physical qubit to that of the encoded qubit, whereas our definition compares the two in a situation where the encoded circuit is expected to operate for significantly longer in order to implement the same logical operation cleanly. This depresses the crossover point to be below error rates for which the error corrected qubit lasts longer than the physical one.
\subsubsection{Selected Logical Error Biases}
\begin{figure*}[t]
    \includegraphics[trim = 100 0 100 0, width = \linewidth]{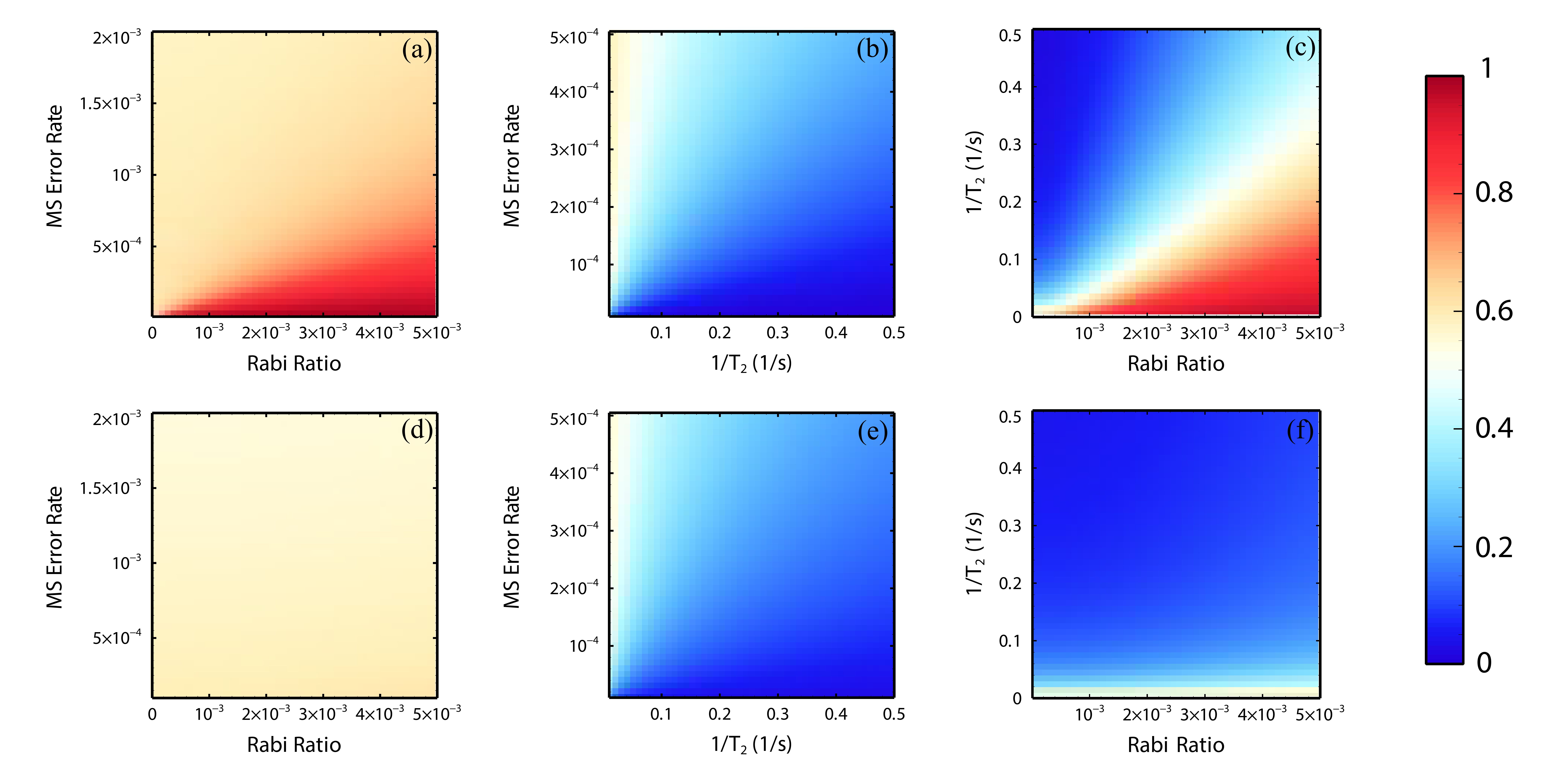}
    \caption{Plots of $Bias_{ZZ}$, which is defined in Equation~\ref{eq:bias} We present data for Bacon-Shor-13 (top row) and Surface-17 (bottom row) over a series of different error models. In the rightmost plots there is also a background overrotation error rate of $p_{2q} = 10\:p_{1q} = 0.0001$.}
    \label{fig:bssfbias}
\end{figure*}
Due to the format of our data, we are able to separately consider the rates at which the $XX$ and $ZZ$ parity is violated, allowing us to consider the bias of error at the logical level due to these asymmetrically structured error models. In this section we will highlight Bacon-Shor-13 and Surface-17, the logical bias plots for both sides of Shor's code can be found in Appendix~\ref{sec:appendixBiases}. The value we show in this plot is 
\begin{equation}\label{eq:bias}
    Bias_{ZZ} = \frac{\langle 1 - Z_{L1}Z_{L2}\rangle}{\langle 1 - X_{L1}X_{L2}\rangle + \langle 1 - Z_{L1}Z_{L2}\rangle}
\end{equation}
where $\langle 1 - Z_{L1}Z_{L2}\rangle$ and $\langle 1 - X_{L1}X_{L2}\rangle$ are the rates at which the $ZZ$ and $XX$ parities are violated.

From the plots in Figure~\ref{fig:bssfbias} we can see that the most strongly biased error is $T_2$ dephasing. While both dephasing and crosstalk errors always appear as a specific type, crosstalk errors during $Z$-type stabilizers are often found sandwiched by single qubit gates which convert them to $Z$-type errors. This factor, along with our crosstalk mitigation techniques preventing dangerous $X$-type logical error pairs from being adjacent in our chains, leads to crosstalk being only slightly biased towards violating $ZZ$ in the cases of Shor-6Z2X and Surface-17. For Bacon-Shor-13 there was not enough freedom for us to implement any of our crosstalk techniques, so the native $X$-type bias is quite strong. In Shor-6X2Z, the larger number of $X$ stabilizers means that there is an asymmetry in the number of error locations in which a crosstalk error would be conjugated into a $Z$-type error, leading to a significant $X$-type bias as well. From these results we can see that the way in which we try to prevent crosstalk through chain ordering has a strong impact on the bias at the logical level. It is possible that by intelligently picking this bias to interface with the underlying error models, we would be able to create an effective error model at the logical level with significant structure.
\section{Conclusion}
In this work we have shown that there are a wide variety of optimal codes when considering different error sources, indicating the importance of being able to accurately benchmark a system and find the error models and parameters which describe it. In the depolarizing error model Surface-17 clearly outperforms all other codes, however when considering physically realistic error models Bacon-Shor-13 and Shor's code variants perform better. Due to the all to all connectivity present in ion trap systems, the surface code is not benefited significantly from its locality. Interestingly, even when considering superconducting systems that have nearest neighbor interactions, other codes continue to outperform the surface code when considering experimental constraints~\cite{chamberland2019topological}.

We also provide evidence for how damaging crosstalk errors really are, further justifying the efforts in looking for methods outside of QECCs for solving it. Implementing a physical or pulse level solution to mitigate the effects of crosstalk will be vital in allowing us to consider a wider variety of codes on our systems. If crosstalk can be lessened or reduced coherently, Bacon-Shor-13 seems to be well suited to solving the other errors present in ion trap systems. We also see that by making choices about how our chains are ordered, we can affect the logical error biases, which could be used to make a more optimized asymmetric code in the future.
\section{Acknowledgements}
The authors thank Michael Newman and Leonardo Andreta de Castro for helpful conversations.  This work was supported by the Office of the Director of National Intelligence - Intelligence Advanced Research Projects Activity through ARO contract W911NF-16-1-0082, National Science Foundation Expeditions in Computing award 1730104, and National Science Foundation STAQ project Phy-1818914.
 
% \clearpage
\bibliographystyle{apsrev}
\bibliography{References}
\onecolumngrid
\appendix
\clearpage
\begin{figure}[t]
    \includegraphics[trim = 100 0 100 0, width = \linewidth]{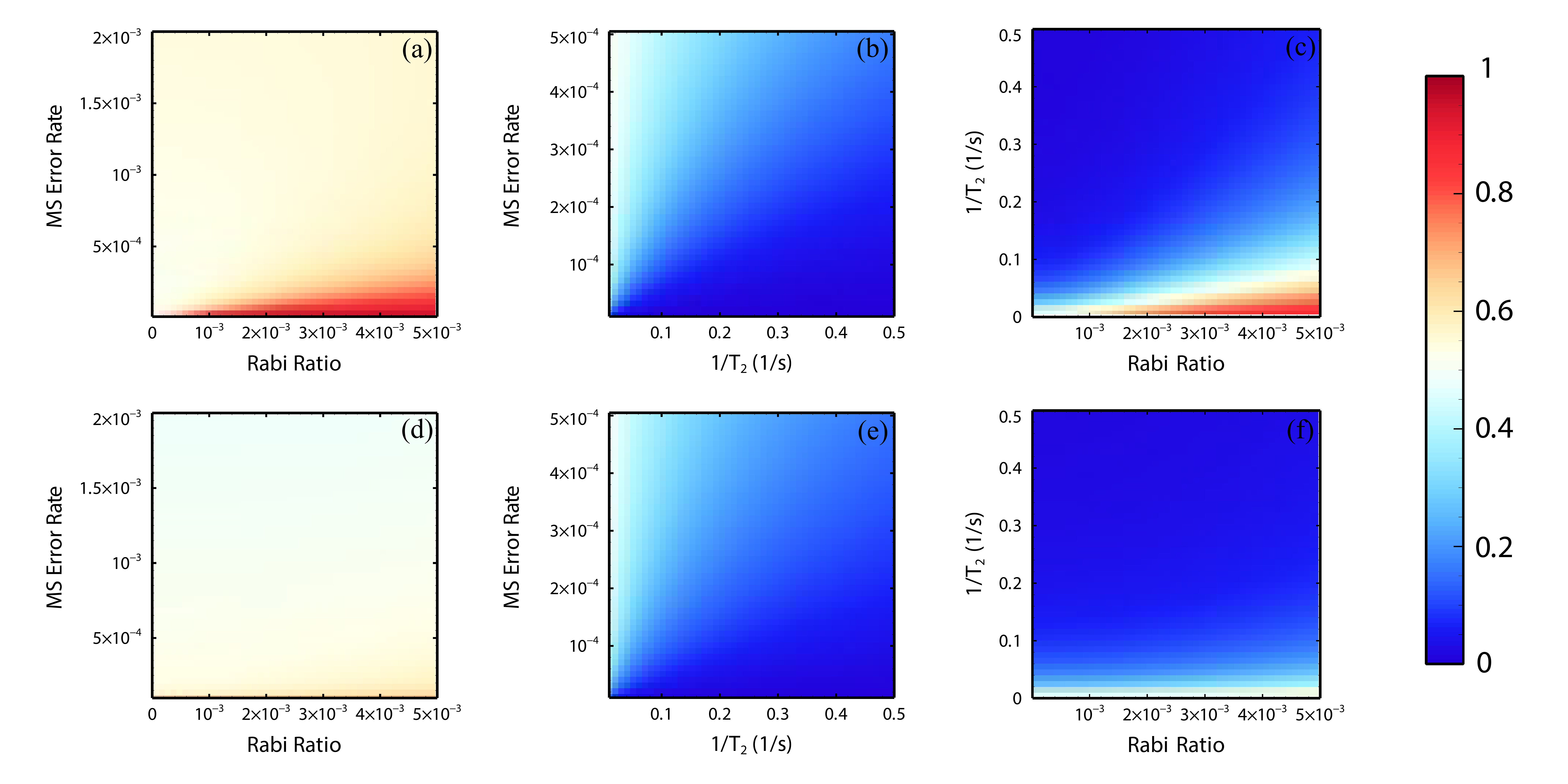}
    \caption{Plots of $Bias_{ZZ}$, which is defined in Equation~\ref{eq:bias} We present data for Shor-6X2Z (top row) and Shor-6Z2X (bottom row) over a series of different error models. In the rightmost plots there is also a background overrotation error rate of $p_{2q} = 10\:p_{1q} = 0.0001$.}
    \label{fig:shorbias}
\end{figure}
\section{Dynamic Programming}\label{sec:appendixDP}
A standard approach for finding the Hamiltonian path of a graph $G=(V,E)$ is dynamic programming~\cite{bellman1961dynamic}. In this method, one determines whether there exists, for each subset $S\subset V$ of vertices and each vertex $v\in S$, a Hamiltonian path that covers $S$ and ends at $v$. For each $(S,v)$, a path exists if and only if a path exists for $(S-\{v\},w)$ for some $w \in S-\{v\}$ such that $(v,w)\in E$. Note that one can look up already-computed answers to avoid redundant computation. Since there are only $O(n2^n)$ number of choices of $(S,v)$, and enumerating $w \in S-\{v\}$ takes $O(|V|)$ time, the total time complexity is $O(|V|^2 2^{|V|})$.

For our problem, it is possible that an ion chain without bad crosstalk does not exist. In this case the problem becomes finding the path that touches each vertex once and requires the fewest additional edges added to the graph. To use dynamic programming, one can ask the following question instead: for each subset $S\subset V$ of vertices, $v \in S$, and non-negative integer $n$, does there exist a path that covers $S$ while touching each vertex once, ends at $v$, and only requires $n$ extra edges added to the graph. For each tuple $(S,v,n)$, a solution exists if and only if one of the following two cases happens:
\begin{enumerate}
    \item Solutions exist for $(S-\{v\},w,n-1)$ for some $w\in S-\{v\}$, however $(v,w) \notin E$. As a result an edge must be added which introduces a bad crosstalk pair.
    \item Solutions exist for $(S-\{v\},w,n)$ for some $w\in S-\{v\}$ such that $(v,w)\in E$, then the edge that is added does not introduce any bad crosstalk pairs.
\end{enumerate}
$n$ has a trivial upper bound $|V|$ since we can definitely use $|V|$ paths to cover the vertex set. Therefore the time complexity is $O(|V|^32^{|V|})$. 
\section{Shor's Codes Logical Bias Plots}\label{sec:appendixBiases}
In Figure~\ref{fig:shorbias} we see the same data as shown in Figure~\ref{fig:bssfbias} but for Shor-6X2Z and Shor-6Z2X. The only major difference between these two codes is that Shor-6X2Z has crosstalk that preserves its bias towards $X$, while Shor-6Z2X is more neutral. They both also have a slightly more $Z$-biased error rate in the overrotation vs. crosstalk plots. As mentioned in the main text, the discrepancy in bias for crosstalk can be explained by the amount of time in the circuit during which a crosstalk error would end up being conjugated into a $Z$-type error, along with the lower number of $X$-type logical operators present for Shor-6Z2X.
\end{document}